# The Model of Magnetic-Field Generation with Screw Dynamo

Andrey G. Tlatov

*Kislovodsk mountain astronomical station of the Pulkovo Observatory, Russian*
tlatov@mai.ru

This paper considers a possibility of magnetic-field generation by local turbulent flows at the bottom of convective zone. The cycle of magnetic-field generation in this model can be represented in the form of sequency of processes. There are vortexes with azimuth axis, similar with Taylor vortex, close to the bottom of convection zone. This leads to the generation of twisted flux tubes because of screw dynamo. The growth of magnetic field causes emersion of U- loops. During the process of emersion and extraction azimuthal field of flux tubes converts to axial field, and reaches the surface as bipolar of sunspots with U-shaped configuration. Due to differential rotation residual bipolar fields stretch out to the surface toroidal field and are shifted to the bottom of the convective zone by means of meridional flow at high latitudes. The direction of the toroidal field within the generation zone reverses its sign, and the cycle is repeated.

A new prognostic index of spots activity is offered on the basis of this model. Prognosis of the following cycle, according to the data of characteristics of sunspots in the current cycle, has rather high correlation coefficient R=0,88.

*Keywords: Solar cycle ; Dynamo model;*

## 1. Introduction

Dynamo action is generally believed to be at the origin of the magnetic field in most astrophysical objects. Nowadays it is considered that generation of the magnetic field of the Sun occurs due to differential rotation. As a result, a toroidal field is formed out of poloidal magnetic field, and it is concentrated in spherical shell at the bottom of convective zone. At the same time, the fields under observation on photosphere are presented in the form of local magnetic field tubes of relatively small size (Schussler, 2004). Another problem is that the value of primary poloidal magnetic field works out several gauss, and the intensity of the observed fields of the sunspots makes several kilogauss. There are reasons to think that the strength of magnetic field is even by several orders of magnitude greater in the generation zone. However, due to ω-effect during the cycle the magnification coefficient is not exceeded by one or two orders of magnitude (Schussler, 2004; Brandenburg, 2005).

To solve these problems, a generation field model different from ω-effect is presented. This model considers the possibility of the generation of the magnetic field by local eddy currents at the bottom of the convective zone.

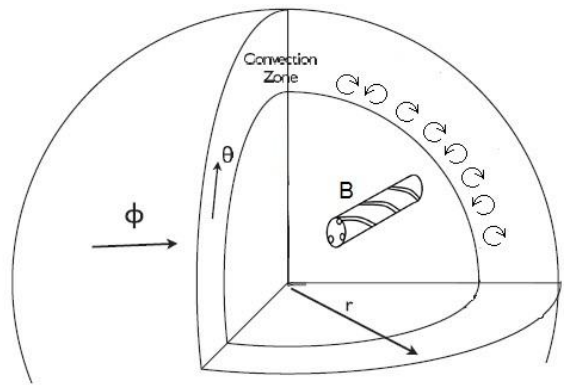

**Fig. 1**. Diagram of convective vortexes at the bottom of the convective zone participating in the process of generation of magnetic field.

## Description of the Model

One of the basic contradictions of the existing paradigm of dynamo mechanism is the difficulty of explaining the features of magnetic bipoles that manifest themselves as sunspots, as a result of emersion of the parts of toroidal magnetic field, which was formed at the bottom of the convective zone in the shape of "a list" with intensity value about $B \sim 10^5$ Gs.

To find the solution we can assume, that generation of the magnetic field tubes represents local processes, in accordance with the mechanism different from ω-effect. This mechanism can be performed by convection cells, existing in the tachocline region. Formation of such cells occurs during the interaction between convective transport and Coriolis force, inflecting ascendant flows towards the poles, and descending flows – towards the equator. Such vortex motions were received by means of numerical simulation of flows at the bottom of the convective zone of the Sun (Rogers & Glatzmaier, 2005). The possibility of generation by vortex motions is well known because of the conducted experiments (Frick et al., 2002).

The cycle of magnetic field generation can be performed by a sequence of processes in this model (Fig.2). The stream of meridional circulation transfers poloidal magnetic field from surface layers to the bottom of the convective zone (1). Nearby the solar tachocline there are convection cells, which lead to twisting of toroidal field that has a configuration of "a list", and generation of local magnetic flux tubes with a screw magnetic field (2). Under the influence of emersion the twisted flux tubes lose their stability and rise to the surface (3). As a result, detachment and magnetic field reconnection take place, circular structures are formed, where magnetic field has a configuration of a twisted field tube (4). Due to emersion and extension of circular structures, their azimuth magnetic field converts into a field directed along an axis. When surfacing to the photosphere, these tubes are observed as sunspots. Under the photosphere the tubes have a U-configuration (5). Magnetic field of the tubes is stretched by differential rotation, thus forming surface toroidal field, which is transferred by meridional circulation to the poles and the zone of generation (6). Further, the cycle is repeated.

The current model is based on four new hypotheses: 1) generation occurs by means of local vortex flows at the bottom of the convective zone; 2) conversion of azimuth field into axial takes place when stretching ring twisted flux tubes during ascent to the surface; 3) Magnetic field of U-tubes is converted into a surface toroidal field due to differential rotation; 4) meridianal transportation shifts toroidal field from the surface to the zone of generation.

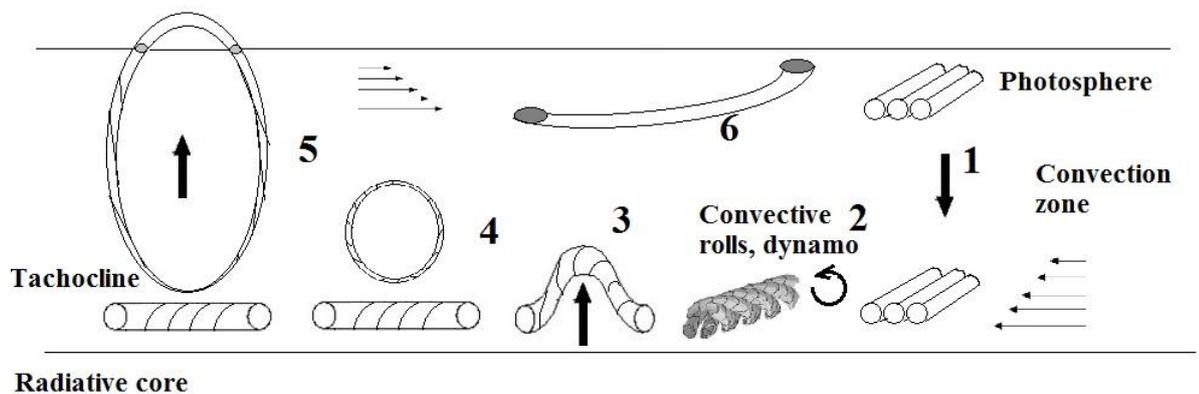

**Fig. 2**. Diagrammatic illustration of solar magnetism with the involvement of screw dynamo.

## Testing of the Hypothesis

A complete testing of the hypothesis involves information concerning many parameters, for instance, rotation rate of vortexes, their size, efficiency of conversion azimuth magnetic field into axial one etc. However, a sufficient difference of the current hypothesis from other existing models is that strengthening of the field takes place in local areas of the existing vortex. And it requires the condition when twisted tubes of limited longitude extent surfaced as ring-type structures. Then, they form toroidally-directed field at the surface, which is the base for the generation of the following cycle. Therefore, we can qualitatively test this hypothesis.

Intensity of the following activity cycle will be proportional to intensity of the current cycle and time, during which stretching of remaining U-tubes into new surface toroidal magnetic field takes place (stage 6, Fig. 2). The lower the altitude of emersion of the spots, the longer is the time period.

Let's make the following calculations. Total area of sunspots per cycle $\sum S_{sp}$ is multiplied by weight-average altitude of the emerged sunspot groups, reckoned from poles $<\theta>$. In this case, the groups will be counted not throughout the whole period of existence of the group, but once, at the moment of its largest area. It turns out, that the value $<\theta>$ in this case is minimal for the 19$^{th}$ cycle and is equal to 72$^o$.

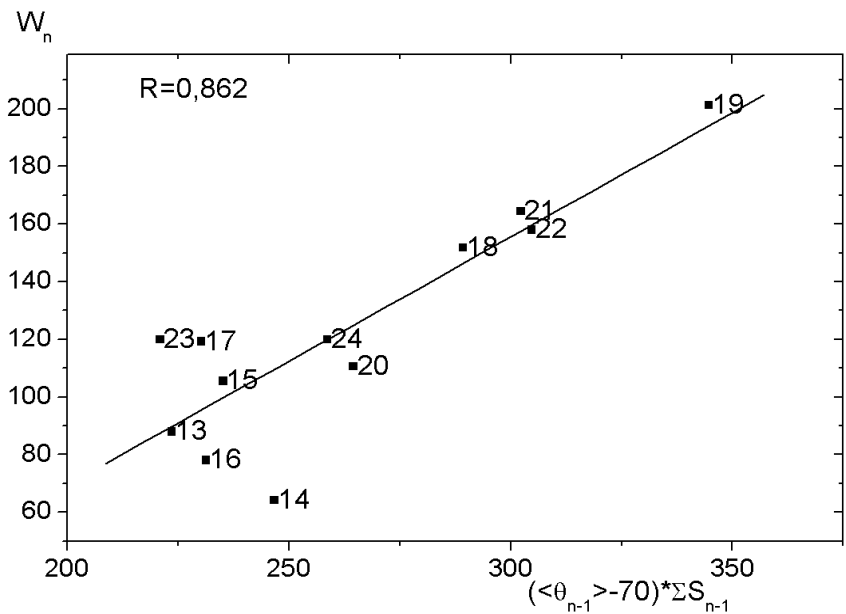

**Fig. 3**. Correlation of parameters of the previous cycles n-1 with the amplitude of the following activity cycle n, owing to stretching of U-bipoles into a surface toroidal field.

## Discussion

The performed above model of generation of the solar magnetic field allows to explain such effects as generation of strong fields of sunspots, observed twisting of magnetic field tubes, accounts for processes of disappearing of the spots. This hypothesis can be checked by means of observations. In particular: 1) there must be groups of sunspots with different direction of twisting of magnetic field. 2) Groups of sunspots which emerged not far from each other can have different directions of twisting.

At the same time, testing of the hypothesis, presented on Fig. 3, gives reasonable grounds for further development of this model.